\documentclass{article}
\usepackage{graphicx} 
\usepackage[utf8]{inputenc}
\usepackage{cite}

\title{STA_RX Paper}
\author{
  Thenmozhi Elango (Department of Electrical Engineering, \\
  Anna University, Chennai, Tamilnadu,India) \\
  \and
 Samiyuktha Kalalii (Department of Electrical Engineering, \\ 
 Indian Institute of Technology Hyderabad) \\
  \and
  P. Rajalakshmi
   (Department of Electrical Engineering, \\ 
 Indian Institute of Technology Hyderabad) 
}
\date{Aug 5, 2014}

\begin{document}
\title{A Novel Approach to Ultrasound Beamforming using Synthetic Transmit Aperture with Low Complexity and High SNR for Medical Imaging}
\maketitle

\section{Abstract}
This paper presents an architecture for Ultrasound Beamforming using Synthetic Transmit Aperture with Low Complexity and High SNR for medical imaging. Synthetic
Transmit Aperture is a novel approach in ultrasound imaging
system by which frame rate and image quality is increased
significantly on less data-transfer and computational requirements. The real-time beamforming performance of Phased
Array(PA) method is limited by high computation and cost.
Thus STA method(data-transfer rate-8MB/frame) advances
over the Phased Array Method(data-transfer rate-95MB/frame)
with comparitively much higher frame rate and Signal to
Noise Ratio(SNR. In this paper, we have implemented receive
beamforming using Synthetic Transmit Aperture (STA) method
for eight channels and have obtained the sample data for
reconstruction of image. The experimental results are compared with the conventional phased array and linear array
beamforming, where it can be observed that the reduction in
memory requirement and high SNR.

\section {INTRODUCTION}

Ultrasound imaging is a technique that has become much
more prevalent than other medical imaging techniques
since this technique is more accessible, less expensive,
safe, simpler to use and produces images in real-time.
However, images produced by an ultrasound imaging
system, must be of sufficient quality to provide accurate
clinical interpretation. The most commonly used image
quality measures are spatial solution, image contrast and
frame rate. In conventional ultrasound imaging system,
when one transducer is used, the quality of images directly
depends on the transducer acoustic field.

Also in conventional ultrasound imaging the image is
acquired sequentially one image line at a time that puts a
strict limit on the frame rate that is important in real-time
imaging system. Thus the moving structures are not easily
imaged and diagnosis may be impaired. This limitation
will seriously halt the image reconstruction when going for
portable devices. An alternative way to obtain an appropriate
spatial resolution, without decreasing of the frame rate, is to use the Synthetic Transmit Aperture Technique \cite{trots2010}.

The basic idea of the STA method is to combine
information from missions close to each other. It provides
the full dynamic focusing, both in transmit and receive
modes, yielding the highest imaging quality\cite{nikolov2010}. In the
STA method at each time one array element transmits a
pulse and all elements receive the echo signals. The data
acquired simultaneously from all directions over a number
of emissions, and the full image can be reconstructed from
these data. The advantage of this approach is that a full
dynamic focusing can be applied to the transmission and
the receiving, giving the highest quality of image \cite{gammelmark2003}.

\begin{figure}[h]  
    \centering
    \includegraphics[width=0.7\textwidth]{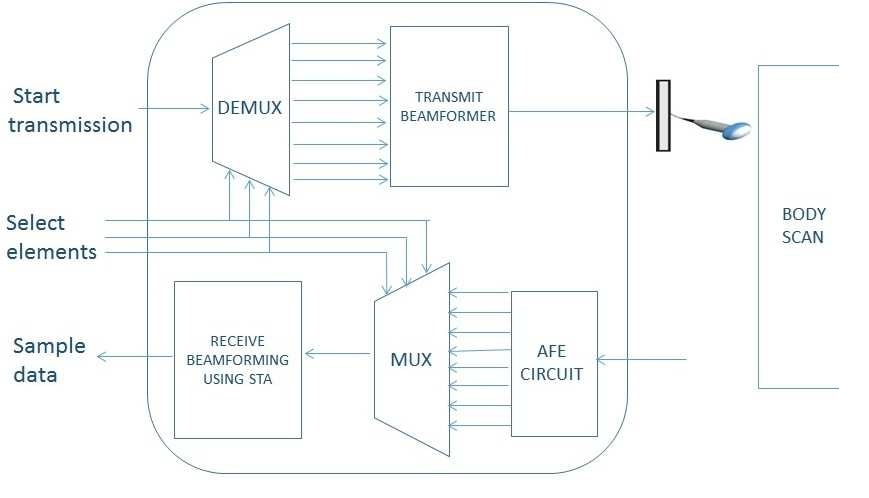}  
    \caption{Architecture of STA ultrasound beamforming}
    \label{fig: image_1.png}
\end{figure}

This is a contrast to the conventional beamforming, where
only imaging along one line in receiving is used \cite{lewandowski2010}. In the
STA method every scan line is imaged as many times as
the number of elements used. This will create an equal
amount of low resolution images which are summed up
to create one high resolution image.The performance of
a system is directly proportional to the size of the active
aperture, which translates into a large number of (primarily
receive) channels. The cost of the system is also directly
proportional to the number of channels. Ideally we would
like to create a large imaging aperture without an increase
in the number of processing channels for memory concerns.

Current systems increase the number of active elements
in the beam former and better engineering makes it possible
to increase the transducer center frequency for the same
penetration depth, which lowers the frame rate, if the image
quality has to be maintained.When comes for flow estimation
the problem is still increased, since several pulse-echo lines
have to be used per estimate and this correspondingly lowers
the frame rate. Also, conventional beam forming methods
suffer from impedance mismatch between the transducer
and receiver circuit which results in low SNR. This can
be overcome by exciting multi element sub aperture method.

A further problem in conventional imaging is the single
transmit focus, so that the imaging is only optimally focused
at one depth. The transmit focusing is thus synthesized
by combining the low resolution images, and the focusing
calculation makes the transmit focus dynamic for all points
in the image. the focus is therefore, both dynamic in
transmit and receive and the highest possible resolution for delay-sum beam forming is obtained everywhere in the
image.

For single-element aperture, the transmit aperture is split
into N subapertures. Each subaperture transmits a prefocused
beam or non-focused beam and dynamically receives the
echo signals. Therefore, Low Resolution Images(LRIs)
is formed at each transmission, and the High Resolution
Image(HRI) is formed by combining all LRIs when all the
elements have transmitted. For N transmissions, there are N
number of LRIs to form a HRI. Since the HRI of STA is
formed by LRIs, the STA image is susceptible to motions \cite{sakhaei2006}.

The experiments conducted for an eight channel phased
array transducer shows the time required to transmit a
beam and receive the echoes through the focused organ
is comparatively less. The number of sample data for
image reconstruction is also greater when compared to the
conventional beamforming.

\section{PROPOSED ARCHITECTURE FOR STA AND MSTA}
The proposed architecture for Ultrasound Beamforming using Synthetic Transmit Aperture with Low Complexity and High SNR for medical imaging is shown in Fig. 1. From the architecture we infer that, with the selection of number of channels at transmit side, receive beamformingwith STA correspondingly generates the scan line sample data which is further required for image reconstruction.

We designed a reconfigurable low complex and high frame rate Synthetic transmit Aperture method for eight channels, which can be extended to any number of  channels. The main idea of our invention is to reduce the memory requirement for producing the sample data when going for more transducer elements. Also, for portable ultrasound imaging systems it is of high priority to form a clear and excellent image quality for clinical interpretation.

The proposed architecture is designed at a frequency of 20 MHz clock, and an inbuilt delay circuit for the receive beamforming which includes the coarse and fine delays,  selection lines to select the transducer element to activate
the channel. Once the transmit beamforming circuit excites the pulses to form a ultrasound signal through a single element, the receiver circuit will be active to receive the echoes from all the channels of the transducer. Thus it provides a full length data construction for imaging.

\begin{figure}[h]  
    \centering
    \includegraphics[width=0.7\textwidth]{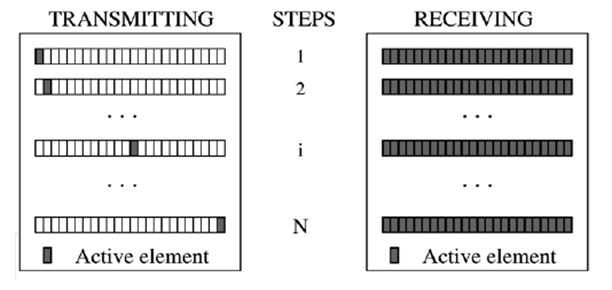}  
    \caption{STA method for single transducer elements}
    \label{fig: image_2.png}
\end{figure}

The proposed architecture can be applicable for all the modes of ultrasound imaging which includes Amplitude mode (A - mode), Brightness mode (B - mode), Colour
mode (C - mode) and motion mode (M- mode). Single Element STA
¯
: In the STA method focusing is
performed by finding the geometric distance from the
transmitting element to the imaging point and back to the
receiving element. The structure of the synthetic aperture
and geometric relation between the transmit and receive
element combination is shown in Fig. 2. When a short
pulse is transmitted by element m and the echo signal is
received by element n, as shown in Fig. 2, a round-trip
delay is T(m,n) = T(m) + T(n), where (m,n) is a transmit
and receive combination \cite{lockwood1998}.
Multi element STA : The major drawback of STA
imaging is the low SNR. Since a single element is used
at each transmission the SNR is very low compared to
linear array imaging, which significantly limits its clinical
application. To overcome this, a subaperture consisting
of multiple grouped elements can be used to emulate a
spherical wave, and hereby increase the SNR. This concept
is called multielement synthetic transmit aperture imaging.
Both groups show that by properly delaying the individual
elements in the subaperture, good approximation to a
spherical wave can be obtained along with a significant
improvement in SNR \cite{jensen2013}.
Applications which involve large penetration depth,
lateral resolution and high SNR (Signal to Noise Ratio)
requires more than one element to generate the ultrasound
signals\cite{hassan2012}. This method of invention is called Multi element
Synthetic Transmit Aperture (MSTA).
In this method small number of elements are used to
transmit a pulse but all array elements receive the echo signals. In practice, it is not very expensive to build a
large transmit aperture, but it is very complex to form
a large receive aperture. For a transmit pulse (from all
transmit subaperture elements), the RF echoes for all receive
elements are stored in memory \cite{shi2011}. When all RF echo
signals have been acquired, the total RF sum is formed by
coherently adding them. The multi-element STA method is
proposed to increase the system frame rate and the speed
of the image acquisition is determined by the number of
transmissions.

\begin{figure}[h]  
    \centering
    \includegraphics[width=0.7\textwidth]{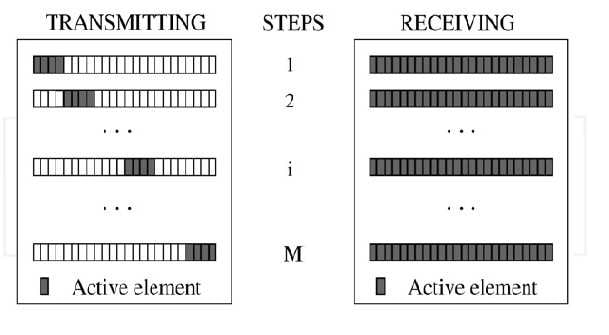}  
    \caption{Multi element STA}
    \label{fig: image_3.png}
\end{figure}

The geometrical locations of the transmit elements in a
multi-element array system impacts the radiation pattern of
that system, which in its turn impacts lateral resolution of
the image, whereas the number of active transmit elements
directly influences the transmitted energy and the SNR.
These parameters define the ultrasound image quality.
Therefore, the optimization of a multi-element STA imaging
system can be formulated as an optimization problem of
the location and the number of the transmit elements.
The main optimization criterion in the multi-element STA
method is the minimal width of the main-lobe combined
with the minimum side-lobe level\cite{ylitalo1994}. This optimization
leads to increasing image penetration depth and high lateral
resolution \cite{jensen2006}.

\begin{figure}[h]  
    \centering
    \includegraphics[width=0.7\textwidth]{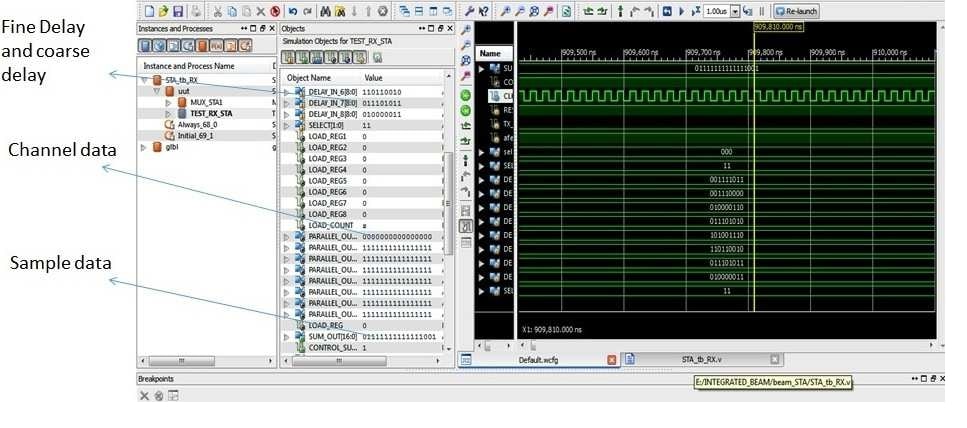}  
    \caption{Simulation results for STA Ultrasound for eight channels}
    \label{fig: image_4.png}
\end{figure}

Depending upon the application and penetration depth
to form a focal point, our design provides an accessibility
for performing STA and MSTA through the selection
elements. This correspondingly reduces the time taken
to form frame rate for imaging, utilizing less memory
resources. Near field and Far field applications such as
cardio vascular applications requires high lateral resolution
and penetration depths provides less accurate frame rates
for image reconstruction. This can be overcome with the
MSTA method of invention.

\begin{figure}[h]  
    \centering
    \includegraphics[width=0.7\textwidth]{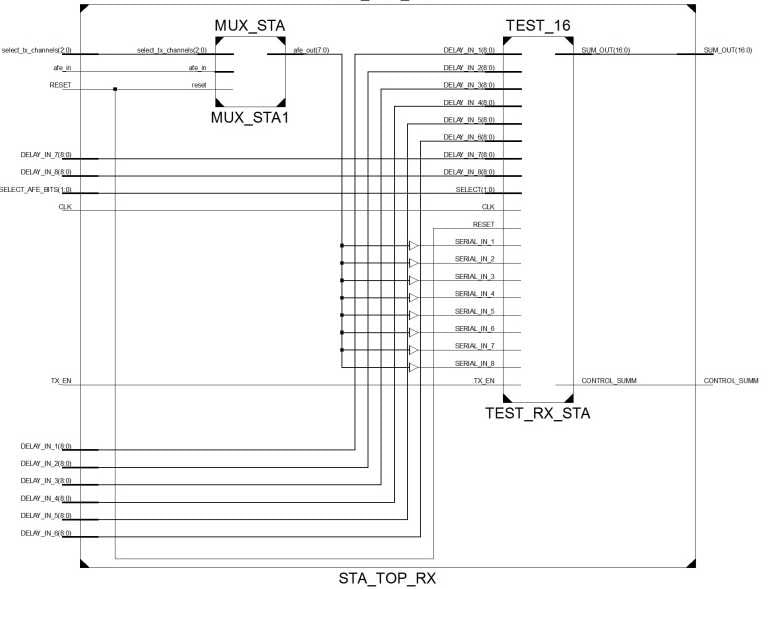}  
    \caption{Synthesized architecture for STA receive beamforming}
    \label{fig: image_5.png}
\end{figure}

\section {EXPERIMENTAL RESULTS}
The proposed dessign for Ultrasound beamforming with
STA is developed using verilog HDL on Xilinx ISE 14.4
tool. Fig. 4 shows the simulation results for the proposed architecture designed for eight channels at a clock frequency of 20 MHz. Fig. 5 shows the synthesized RTL implementation
generated through Xilinx ISE 14.4 tool. The obtained results
show that the design can be easily configured into hardware
for practical applications.

\section{CONCLUSION}

In this paper we proposed the design of ultrasound
beamforming with Synthetic transmit aperture method. The
proposed architecture utilizes less memory with increased
frame rate for medical imaging purposes.

\newpage
\bibliographystyle{plain}
\bibliography{reference}

\end{document}